\documentclass[twocolumn,showpacs, showkeys, pra]{revtex4}
\usepackage{subfigure}
\usepackage[pdftex]{graphicx}
\usepackage{epsfig}

\begin{document}

\title{Dissipative Nonlinear Josephson Junction of Optical Soliton and Surface Plasmon}

\author{Yasa Ek\c{s}io\u{g}lu} 
\email{yeksioglu@ku.edu.tr}
\author{\"{O}zg\"{u}r E. M\"{u}stecapl{\i}o\u{g}lu}
\author{Kaan G\"{u}ven}
\affiliation{Department of Physics, Ko\c{c} University, Istanbul 34450\ Turkey}
\date{\today}

\begin{abstract}
We examine the dynamics of a dissipative photonic Josephson junction formed by the weak coupling of an optical soliton in a nonlinear dielectric waveguide and a co-propagating surface plasmon along a parallel metal surface with a linear dielectric spacer. We employ a heuristic model with a coupling function that depends on the soliton amplitude, and consider two phenomenological dissipation mechanisms separately: angular velocity dissipation and population imbalance dissipation. In the former dissipation mechanism, the system exhibits phase-slip phenomenon where the odd-$\pi$ phase modes decay into even-$\pi$ phase modes. The latter damping mechanism sculptures the phase-space significantly by introducing complex features, among which Hopf type bifurcations are notable. We show that some of the bifurcation points expand to stable limit cycles for certain regimes of the model parameters.
\end{abstract}

\maketitle
\section{Introduction}
\label{sec:Introduction}

As the research on the interaction of light with metallic structures matures as a well-established technology called plasmonics~\cite{zayats,maier,self}, the coupling of surface plasmons to different light sources are being investigated, with the motivation that controlling surface plasmons offer the potential for developing different types of SP-integrated nanophotonic devices~\cite{surface-plasmon-nanophotonics,ritchie,ybliokh}. In particular, the coupling between SPs and confined light modes are widely investigated.
A recent proposal is based on the resonant interaction between the SPs on a metal surface and co-propagating soliton in a nonlinear dielectric medium~\cite{proposal1}. Under a classical formulation, this system exhibits rich nonlinear dynamical features where the interaction depends on the soliton amplitude, as such it may be utilized to manipulate the SP propagation. As an additional feature, it has been shown~\cite{dynamical} that the system is akin to the bosonic Josephson-junction of Bose-Einstein condensates~\cite{dynamical,proposal10,smerzi97,smerzi,raghavan} so that similar and different nonlinear Josephson junction features may be realized in this optic-plasmonic system.\\
While this was an exciting analogy, surface plasmons are subject to strong dissipative effects in the host metal and perfect Josephson junction is a too idealized model. In the present work, we aim to make a more faithful representation of the system by taking into account the dissipation. Remarkably, we find that the dissipation can bring some benefits if it could be introduced under control. The junction can exhibit stable dynamics in spite of dissipation. Based on this motivation, we investigate here the dissipation effects of the coupled soliton-SP system based on the model introduced in Ref.~\cite{proposal1} and then formulated as a nonlinear Josephson junction~\cite{dynamical}. We note that dissipation effects in the Bosonic Josephson Junction (BJJ) were previously studied,~\cite{proposal10} which might be useful to refer for comparison. We employ a standard dynamical analysis of the system in the phase space representation, and investigate how different dissipation mechanisms and other model parameters affect the phase-space landscape. Bifurcations in the system are highlighted which may be induced in certain regimes of these parameters.
\section{Model and Theoretical Formulation}
\label{sec:Theory}
Construction of a classical theoretical model for a coupled soliton surface-plasmon system has been addressed recently in a number of publications. An elegant initiative was the heuristic model introduced by Bliokh and collaborators,~\cite{proposal1} in which the interaction between the soliton and the surface-plasmon is formulated as a coupled non-linear/linear oscillator system, with the coupling parameter being dependent on the soliton amplitude. While being constrained by a number of simplifying assumptions, the model predicts stationary coupled modes of the soliton and the surface-plasmons under feasible parameter regimes. In a following work, stationary and quasi-stationary solutions for the so-called soliplasmons are investigated under an asymmetric coupling model~\cite{milian_arxiv1205.3182}.
In this paper, we will employ the heuristic coupled-oscillator model, albeit with an improved coupling parameter adapted from the recent formulations. Our main motivation is to incorporate the correct behavior for the coupling in the strong soliton and strong surface-plasmon amplitudes, respectively~\cite{milian_arxiv1205.3182}.
We begin by recapitulating the heuristic theoretical model in Ref.~\cite{proposal1}, where the interaction between optical solitons and surface plasmons (SP) is discussed in the context of an resonant optical system: a soliton propagating parallel to a metal interface can excite surface-plasmons by its evanescent lateral tail, which, in turn, interacts in resonance with the soliton along the propagation. The soliton and the SP are assumed to be co-propagating, hence only the spatial dynamics of the propagation is of concern. The nonlinearity of the system is assumed to be confined at a distance, $d$, from the metal interface such that the surface-plasmon propagation retains linearity, and the coupling between the soliton and SP-fields is weak. The formulation is based on a 2D coordinate system in which $y$ is the propagation direction and  $x$ is the lateral direction.

The total electric field of the system is introduced by the following variational ansatz in which the soliton and SP fields are written in a product form of their respective longitudinal ($c_{p,s}(y)$) and transverse amplitudes. 
 
\begin{equation}
\ E(x,y) = c_p(y)e^{-\kappa_px} + \frac{c_s(y)}{\cosh\left[\kappa_s(x-d)\right]},
\label{eq:Total Field}
\end{equation}
Evanescent wavevectors are $\kappa_p=\sqrt{k_p^2-k^2}$ and $\kappa_s=k\sqrt{\gamma/2}|c_s|$. The $k$ is the incoming wave vector whereas $k_p$ is the SP wave vector. $\gamma$ is the nonlinearity parameter of the dielectric strip. The longitudinal amplitudes $c_{p,s}(y)$ obey the coupled spatial propagation wave equations:
\begin{equation}
\begin{array}{ll}
\ \ddot{c}_{p} + \beta_{p}^2c_{p} = q(|c_s|)c_{s},
\ \ddot{c}_{s} + \beta_{s}^2c_{s} = q(|c_s|)c_{p}.
\end{array}
\label{eq:CoupledOscillator}
\end{equation}
where $\ddot{c}_{p,s}\equiv \frac{\partial^{2}c_{p,s}}{\partial\xi^{2}}$ and $\xi\equiv ky$ is the dimensionless propagation coordinate. $\beta_p = k_p/k$ and $\beta_s=1+\gamma|c_s|^2/4$ are the propagation constants of SP and soliton respectively. $q(|c_s|)$ is the coupling function which we will discuss further below. The coupled equations are linearized by introducing $c_{p,s} = \mathcal{C}_{p,s}e^{i\xi}$ into Eq.~\ref{eq:CoupledOscillator} and by employing the slowly varying amplitude approximation to discard higher order derivatives. The final set of equations are then:
\begin{equation}
\-i\dot{\mathcal{C}}_{p}=\nu_p\mathcal{C}_{p}-\frac{q(|\mathcal{C}_{s}|)}{2}\mathcal{C}_{p},
\label{eq:Cp_and_Cs_1}
\end{equation}
\begin{equation}
\-i\dot{\mathcal{C}}_{s}=-\frac{q(|\mathcal{C}_{s}|))}{2}\mathcal{C}_{p}+\nu_s(|\mathcal{C}_{s}|)\mathcal{C}_{s}.
\label{eq:Cp_and_Cs}
\end{equation}
$\nu_p\equiv\beta_p-1\ll1$, $\nu_s\equiv\beta_s-1\ll1$ are the small deviations of the dimensionless propagation constants of surface-plasmon and soliton, respectively.

Evidently, the coupling function plays the main role in these equations. Qualitatively, it should describe the interaction through the overlap of the evanescent tails of the soliton and plasmon in the (linear) dielectric spacer between the metal interface and the nonlinear dielectric waveguide. Since the lateral profile of a soliton depends on its amplitude, the coupling function is expected to exhibit this dependence, too. An analytical formulation of the coupling function was provided in Ref.~\cite{milian_arxiv1205.3182} with the remark that the equation system is not symmetric in the coupling function in general. Within the heuristic model, the following functional form captures the expected behavior in the strong soliton and strong SP amplitude limits:

\begin{equation}
\label{eq:q(Cs)}
q(|\mathcal{C}_{s}|)\simeq|\mathcal{C}_{s}|e^{-\sigma kd\sqrt{\frac{\gamma}{2}}|\mathcal{C}_{s}|}   
\end{equation}

Here, $\sigma$ is an overall constant which we introduce as a parameter to account for the other constants (e.g. permittivities, propagation constants). The key parameters are preserved in the functional form. In the strong soliton amplitudes ($|\mathcal{C}_{s}|\approx 1$), the exponentially decaying term dominates the coupling. In the weak soliton amplitude, the coupling is proportional to $|\mathcal{C}_{s}|$ since the soliton lateral profile in Eq.~\ref{eq:Total Field} would be wider resulting in a larger overlap. 
 
The analogy between this model and the bosonic Josephson junction (BJJ) dynamics~\cite{raghavan,smerzi97} is revealed when we further substitute $\mathcal{C}_{s,p} = C_{s,p}e^{i\phi_{s,p}}$ into Eq.~\ref{eq:Cp_and_Cs} and introduce a new variable set consisting of the fractional population imbalance $Z = (|C_s|^2 - |C_p|^2)/N$ and the relative phase difference between the soliton and the SP $\phi = \phi_s - \phi_p$:
\begin{equation}
\ \dot{Z}=-q(Z)\sqrt{1-Z^2}\sin\phi,
\label{eq:DWSP_JJ_Z}
\end{equation}
\begin{equation}
\ \dot{\phi}= \Lambda Z+\Delta E+\frac{q(Z)Z}{\sqrt{1-Z^2}}\cos\phi,
\label{eq:DWSP_JJ_phi}
\end{equation}
where $\Lambda=\frac{\gamma N}{8}$, is the nonlinearity parameter, and $\Delta E\equiv\Lambda-\nu_p$ parametrizes the asymmetry between the soliton and SP states (similar to the asymmetry between the wells of a double well system). $N = (|C_s|^2+|C_p|^2)$ is constant for the isolated system. The coupling function $q$ takes the form
\begin{equation}
\ q(Z)\simeq\sqrt{\frac{(1+Z)}{2}}e^{-\sigma kd\sqrt{2\Lambda(1+Z)}}.
\label{eq:q}
\end{equation}
We stress that the $Z$ dependence makes an inherently dynamic coupling as opposed to constant (or externally tunable) coupling parameter present in BJJ-systems.
With this formulation, the coupled soliton SP system can be described as a nonlinear Josephson junction.

The range of the model parameters are chosen to be around the resonant coupling regime~\cite{proposal1}. In the calculations presented below, we fix the dimensionless separation parameter $kd=6$, and choose the other parameters between $\Lambda\simeq 0.003 - 0.03$, $\Delta E \simeq  $-0.4$ - $0.4$ $, accordingly. Evidently, using larger $kd$ values would confine the interaction to smaller soliton amplitudes and eventually result in the decoupling of the soliton and the SP completely. Small $kd$ values would not be applicable in the weak coupling approximation.

\begin{figure*}[ht!]
\epsfig{file=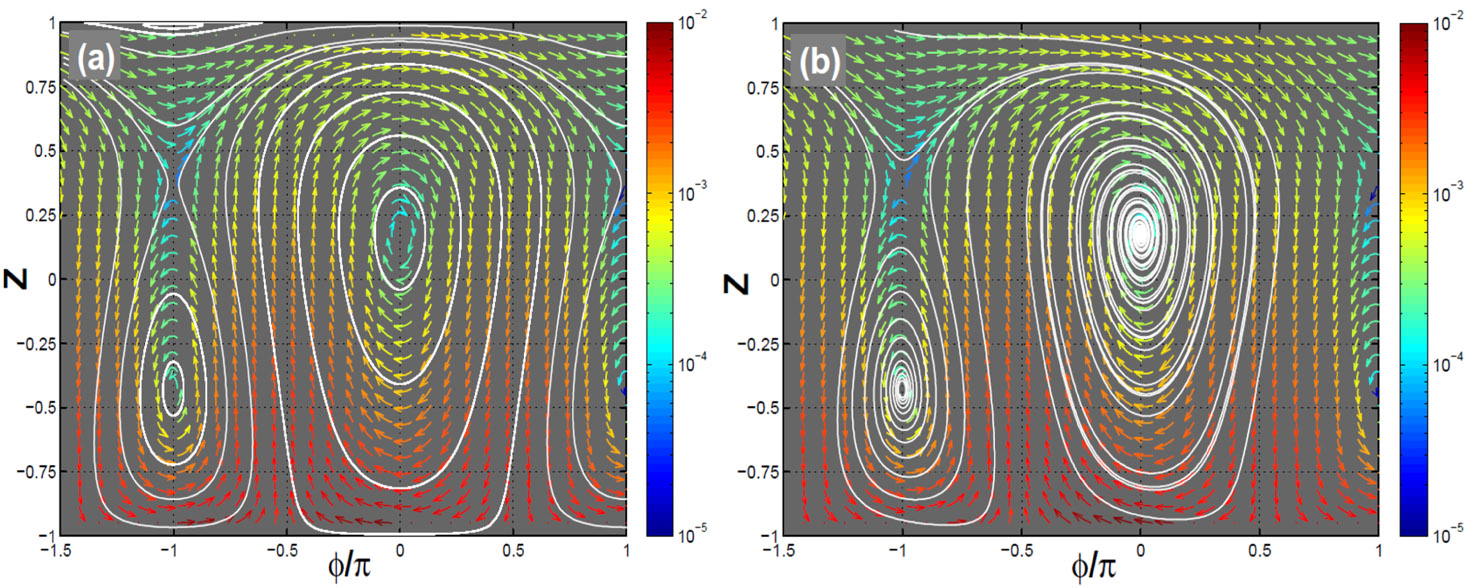,width=0.75\linewidth,clip=}
\caption{\scriptsize (Colour online)(a)The $Z-\phi$ phase-space of the dissipationless SP-soliton coupled system with $kd=6$, $\Lambda=0.01$, $\Delta E= -0.0025$. The magnitude of the gradient plot is color-coded logarithmically. White curves show typical open and closed trajectories, (b)Phase-space trajectory with parameters of (a) and with angular velocity dependent dissipation, $\eta=0.2$. Phase slip occurs between the $\phi=-\pi$ and $\phi=0$ modes.}
\label{fig:1}
\end{figure*}

\section{Angular-velocity dependent dissipation}
\label{sec:phidamping}

Owing to the analogy to the BJJ models~\cite{proposal10}, the first dissipation mechanism we consider is about the incoherent exchange of photons between the soliton and the SP. This is introduced by the term $\eta\dot{\phi}$ in the following equations,

\begin{equation}
\ \dot{Z}=-q(Z)\sqrt{1-Z^2}\sin\phi-\eta\dot{\phi},
\label{eq:DWSP_JJ_Z_phidamp}
\end{equation}
\begin{equation}
\ \dot{\phi}= \Lambda Z+\Delta E+\frac{q(Z)Z}{\sqrt{1-Z^2}}\cos\phi,
\label{eq:DWSP_JJ_phi_phidamp}
\end{equation}

Under coordinate reversal $\xi\rightarrow -\xi$, we have $Z\rightarrow Z$ and $\phi\rightarrow -\phi$, hence, the dissipative term $\eta\dot{\phi}$ appearing in the equation for $\dot{Z}$ represents an irreversable process correctly. The phase-space of the dissipationless and dissipative system are shown in Fig.\ref{fig:1}(a) and \ref{fig:1}(b), respectively. The gradient distribution is given with color-coded magnitudes and typical trajectories are indicated by white curves. The phase space is characterized by the zero-(even) phase and $\pi$-(odd) phase modes, where the critical points are located. There is always one stable point at $\phi=0$. At $\phi=\pi$, one (stable) to three (two stable, one saddle) critical points may emerge depending on the values of the model parameters $\Lambda, kd$, and $\Delta E$~\cite{dynamical}. In Figure~\ref{fig:1}(a), sample trajectories (white curves) show anharmonic closed orbits around stable critical points and open trajectories. The color indicates that the gradients get steeper in the SP-dominant population imbalance region ($Z<0$).

Since the critical points are determined by the $\dot{Z}=0, \dot{\phi}=0$ condition, the $\dot{\phi}$-dependent dissipation does not alter the location of the critical points in the phase space but the eigenvalues of the Jacobian acquire nonzero real parts such that the stable critical points at $\phi=\pi$ becomes source and the critical point at $\phi=0$ becomes a sink. While the trajectories flow towards the sink points, the phase-slip phenomenon is observed from $\phi=$odd-$\pi$-modes to $\phi=$even-$\pi$ modes, as reported for the BJJ systems~\cite{proposal10}. Depending on the value of the asymmetry parameter, $\Delta E$, the phase-slip occurs with a nonzero change in the averaged population imbalance $<Z>$.

With a larger (smaller) dissipation constant, the typical phase-diagram depicted in Fig.\ref{fig:1}(b) gets compressed (extended) along the phase axis but the features described above remain the same. Thus, while the transient behavior depends on $\eta$, the final state of the system is independent: $\phi_{s}=\phi_{p}$ and a fixed value of $Z$.

\begin{figure*}[ht!]
\epsfig{file=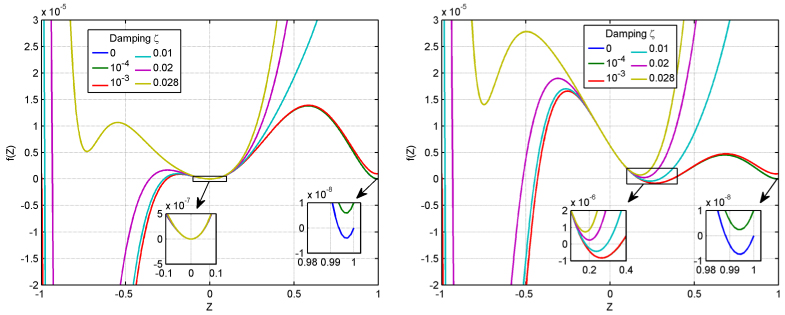,width=0.75\linewidth,clip=}
\caption{\scriptsize (Colour online) (a)Plot of Eq.~\ref{eq:f(z)} for $kd = 6, \Lambda = 0.01$ and $\Delta E = 0$ at various values of the dissipation constant $\zeta$. Roots of $f(Z)$ are the critical points in the phase-space,(b)Plot of Eq.~\ref{eq:f(z)} for $kd = 6, \Lambda = 0.01$ and $\Delta E = -0.0025$ at various values of the dissipation constant $\zeta$. Roots of $f(Z)$ are the critical points in the phase-space.}
\label{fig:2}
\end{figure*}  

\section{Population imbalance dissipation}
\label{sec:zdamping}
In the following, we introduce the set of equations that incorporate a dissipation proportional to the population imbalance, $Z$. In the context of BJJ systems, such a dissipation arises from the finite lifetime of excited states of two condensates in different hyperfine levels in a single harmonic trap connected by tunnelling transitions~\cite{chaohong_lee}. In the analogous mechanical system of momentum shortened pendulum, this corresponds to a damping proportional to the angular momentum variable~\cite{williams}, which we consider to be applicable to the soliton-SP system phenomenologically.

\begin{equation}
\ \dot{Z}=-q(Z)\sqrt{1-Z^2}\sin\phi-\zeta{Z},
\label{eq:Z_damp1}
\end{equation}
\begin{equation}
\ \dot{\phi}= \Lambda Z+\Delta E+\frac{q(Z)Z}{\sqrt{1-Z^2}}\cos\phi.
\label{eq:Z_damp2}
\end{equation}

The first thing we note here is that this dissipation term can modify the critical points in the phase-space significantly: their existence, location and corresponding Jacobian eigenvalues. For determining the critical points, using $\dot{Z}=0, \dot{\phi}=0$ and the identity $\sin^{2}(\phi)+\cos^{2}(\phi)=1$ we obtain the following root equation in $Z$,
\begin{equation}
Z^4\zeta^2+(1-Z^2)[(\Delta E+\Lambda Z)^2(1-Z^2)-q(Z)^2 Z^2]=0
\label{eq:f(z)}
\end{equation}
Figure~\ref{fig:2} is a plot this equation for fixed $kd = 6, \Lambda = 0.01$ and for (a) $\Delta E = 0$, and (b) $\Delta E = -0.0025$, with several order of magnitudes of the dissipation constant $\zeta$.  Blue curve is plotted for dissipationless case ($\zeta=0$) for reference. For $\Delta E = 0$, $Z=0$ is always a critical point as all curves in Fig.~\ref{fig:2}(a) are tangent to $Z=0$ at this point, irrespective of dissipation. This can also be inferred from Eq.~\ref{eq:f(z)}. Critical points close to $Z\approx 1$ are most sensitive to the dissipation strength and they already disappear at weak dissipation ($\zeta\approx10^{-4}$). For $\Delta E = -0.0025$, as the dissipation gets stronger, the critical points collide and annihilate in pairs. This is actually indicating the presence of saddle-node type bifurcations in the system which we show further below. Finally, above $\zeta = 0.0275$ no critical point is present in the phase space. The sensitivity of the critical points to the dissipation strength increases rapidly after $\zeta > 0.01$. Table~\ref{table:damp} classifies the critical points at different dissipation regimes for $\Delta E = -0.0025$. Complex eigenvalues of the Jacobian with positive (negative) real part define spiral source (sink) points and saddle have zero eigenvalues.   

\begin{figure*}[ht!]
\epsfig{file=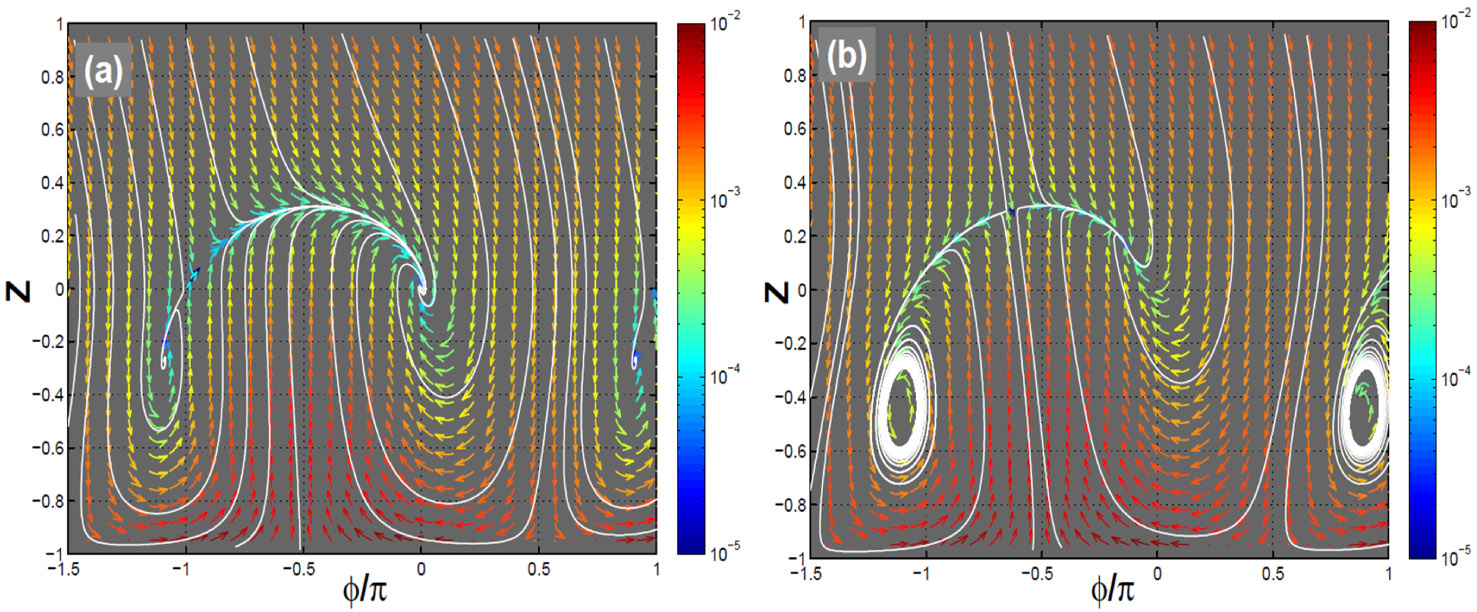,width=0.75\linewidth,clip=}
\caption{\scriptsize (Colour online) Phase-space plot of coupled soliton-SP system with population imbalance dissipation. Model parameters are $kd = 6, \Lambda = 0.01$, $\zeta = 0.01$. (a) $\Delta E = 0$, (b) $\Delta E = -0.0025$. White curves indicate sample trajectories. The magnitude of the gradient plot is color-coded logarithmically. In (b), only trajectories exterior to the limit cycles are shown for clarity.}
\label{fig:3}
\end{figure*}

\begin{figure*}[ht!]
\epsfig{file=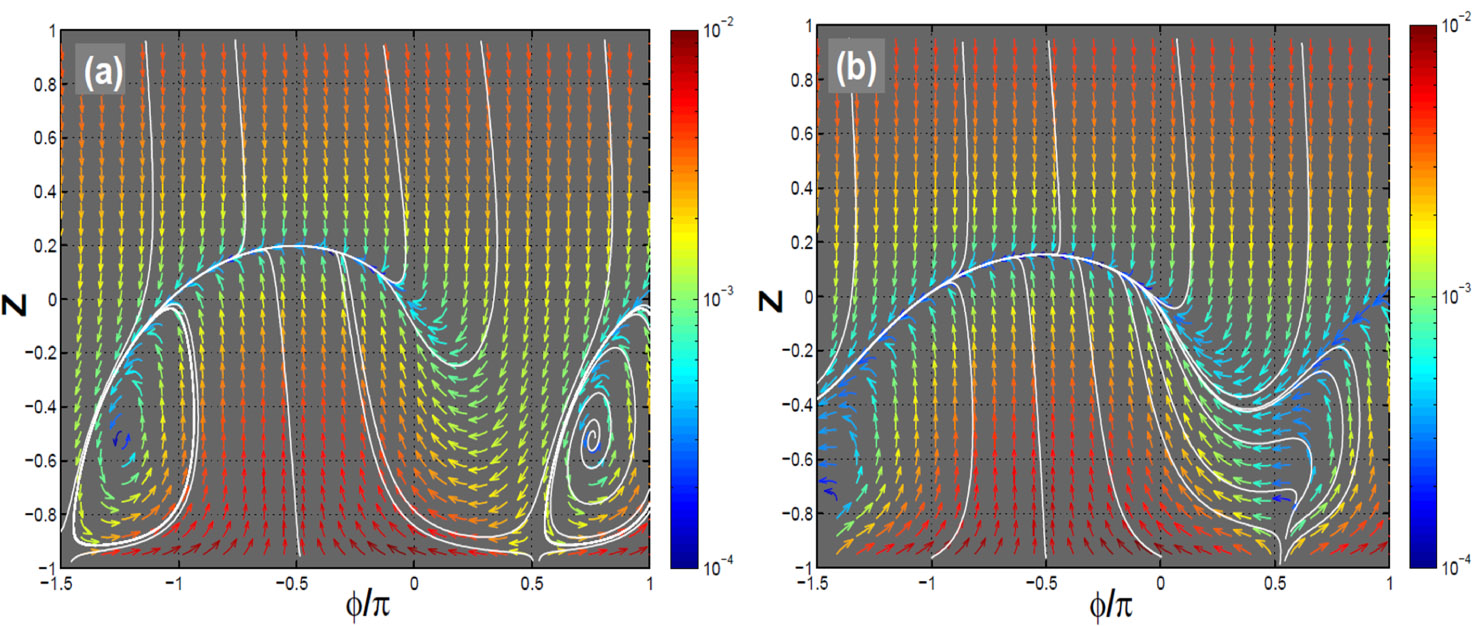,width=0.75\linewidth,clip=}
\caption{\scriptsize (Colour online)Phase-space plot of coupled soliton-SP system with population imbalance dissipation. Model parameters are $kd = 6, \Lambda = 0.01$, $\Delta E = -0.0025$  (a) $\zeta = 0.02$. Exterior and interior trajectories shown for left and right limit cycles, respectively. (b) $\zeta = 0.028$. White curves indicate trajectories. Gradient plot is color-coded logarithmically}
\label{fig:4}
\end{figure*}

\begin{table}[ht]
\caption{Classification of fixed points at different dissipation regimes for Fig.~\ref{fig:2}(b)}
\centering% used for centering table
\begin{tabular}{c c c c c c c} % centered columns (4 columns)
\hline\hline %inserts double horizontal lines
$\zeta$ & I & II & III & IV & V & VI\\ [0.5ex] % inserts table %heading
\hline % inserts single horizontal line
$0$$<$$\zeta$$<$$8.5\times10^{-5}$ & saddle & sso & ssi & saddle & ssi & saddle\\% inserting body of the table
$8.5\times10^{-5}$$\leq$$\zeta$$<$$0.016$ & saddle & sso & ssi & saddle &- & -\\ 
$0.016$$\leq$$\zeta$$<$$0.027$ & saddle & sso & - & - & - & -\\
$0.027$$\leq$$\zeta$ & - & - & - & - & - & - \\ [1ex] % [1ex] adds vertical space
\hline %inserts single line
\end{tabular}\\
{sso: spiral source, ssi:spiral sink}
\label{table:damp} % is used to refer this table in the text
\end{table}
%\newpage
According to the typical ranges of the dissipation parameter we now investigate the behavior of the system in the phase-space. Figure~\ref{fig:3}(a) is plotted for $kd=6$, $\Lambda=0.01$, $\Delta E=0$, and $\zeta = 10^{-2}$. These values corresponds to the cyan curve in Fig.~\ref{fig:2}(b). $Z=0$ is a spiral sink at $\phi=0$ and a saddle point at $\phi=\pi$. Another spiral sink is located at $Z = -0.28, \phi = 0.85\pi$. The saddle point at $Z = -0.98$, $\phi = 0.5\pi$ is not visible at this scale.

In Figure~\ref{fig:3}(b), same parameters as in Fig.~\ref{fig:3}(a) are used but $\Delta E = -0.0025$. Remarkably, a stable limit cycle is present, indicating the existence of bifurcations induced by the model parameters. In Figure~\ref{fig:4}(a) the dissipation constant is increased to $\zeta = 0.02$ for which a larger limit cycle is observable. Finally, $\zeta \approx 0.028$ is the onset of the regime in which all trajectories flow into a unique stable cycle and there are no critical points in the phase space. This behavior is similar to that of a Josephson junction between two superconductors, with bias current above the critical current, $I\equiv\frac{I_{B}}{I_{C}}>1$,(see e.g. Strogatz~\cite{strogatz}). There, the current phase $\phi$ and angular velocity $\dot{\phi}\equiv y$ satisfy the dimensionless equations 
\begin{equation}
\dot{\phi}=y,
\end{equation}
\begin{equation}
\dot{y}=I-\sin(\phi)-\alpha y.
\label{biased-Josephson}
\end{equation}
where $\alpha$ is the damping parameter. Note that for $I>1$, there are no critical points.

\subsection{Codimension-1 Bifurcations}
\begin{figure*}[ht!]
\epsfig{file=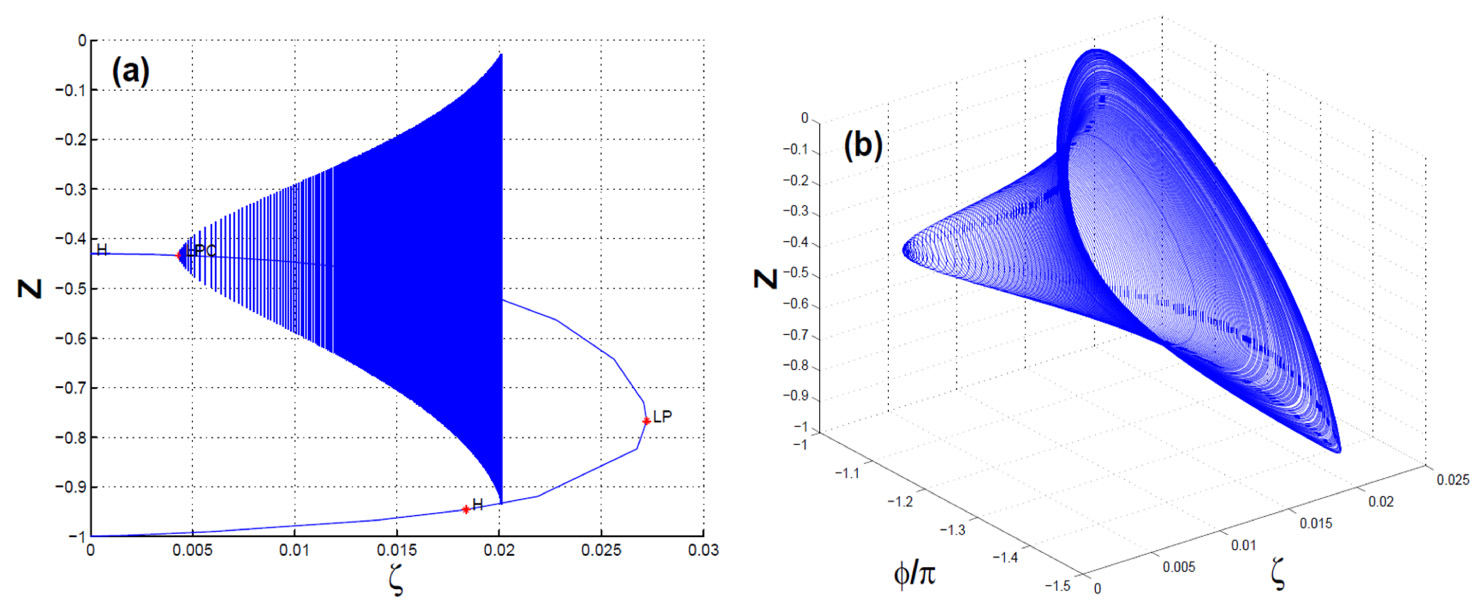,width=0.75\linewidth,clip=}
\caption{\scriptsize (Colour online)(a) Continuation of the stable point as a function of dissipation constant. Model parameters are $kd = 6, \Lambda = 0.01$, $\Delta E = -0.0025$. Andronov-Hopf (H) bifurcation, limit point (LP), and neutral saddle (H) are observed. Vertical lines indicate the extend of the limit cycles along the $Z$-axis. A limit point cycle terminates the limit cycles at $\zeta\approx 2.01$, (b) Limit cycles plotted in the $Z-\phi -\zeta$ parametric phase-space.}
\label{fig:5}
\end{figure*}

In order to investigate the bifurcation dynamics, we first consider codimension one bifurcations which can be induced by varying a single parameter of the model. We employ a numerical continuation software called Matcont~\cite{matcont}, which computes how a critical point advances in the phase space by varying model parameters, and characterizes the point along the continuation according to its Jacobian. We begin by the continuation with respect to the dissipation constant $\zeta$. The fixed parameters are $kd=6$, $\Lambda = 0.01$, and $\Delta E = -0.0025$. Figure~\ref{fig:5}(a) shows the computed continuation curve, on which an Andronov-Hopf (H) bifurcation, limit point (LP), and neutral saddle point (also labeled as H) are observed. These points are characterized with the normal form coefficient in Table~\ref{table:eta_bifurcations}. The Andronov-Hopf bifucation is supercritical, since the first Lyapunov coefficient is negative~\cite{kocak,strogatz,Arrowsmith}. The vertical lines indicate the span of limit cycles in $Z$ axis, emanating from the Hopf-point. Figure~\ref{fig:5} (b) shows the limit cycles in the $Z-\phi-\zeta$ plane. Interestingly, the limit cycles get bigger in the $\phi-Z$ plane for stronger dissipation, before they are terminated by a limit point cycle. For reference, the limit cycles in Figures~\ref{fig:3} and~\ref{fig:4} are in the range of Fig.~\ref{fig:5}(b).

\begin{table}[ht]
\caption{Bifurcation points shown in Fig.~\ref{fig:5}(a)} % title of Table
\centering % used for centering table
\begin{tabular}{l cccc} % centered columns (4 columns)
\hline\hline %inserts double horizontal lines
Label & $\Delta E$& $\phi^{*}$ & $Z^{*}$ & $a$\\ [0.5ex] % inserts table heading
\hline % inserts single horizontal line
H & $0.0437$ & $-1.05$ & $-0.434$ & $-0.0126$\\ % inserting body of the table
LP & $0.0272$ & $-1.42$ & $-0.767$ & $-0.109$\\
NS (H) & $0.0184$ & $-1.48$ & $-0.495$& -\\[1ex] % [1ex] adds vertical space
\hline %inserts single line
\end{tabular}\\
{LP: Limit Point (fold),H:Andronov-Hopf, NS:Neutral Saddle, 
a is the normal form coefficient for LP and first Lyapunov coefficient for H.}  
\label{table:eta_bifurcations} % is used to refer this table in the text
\end{table}

\begin{figure*}[ht!]
\epsfig{file=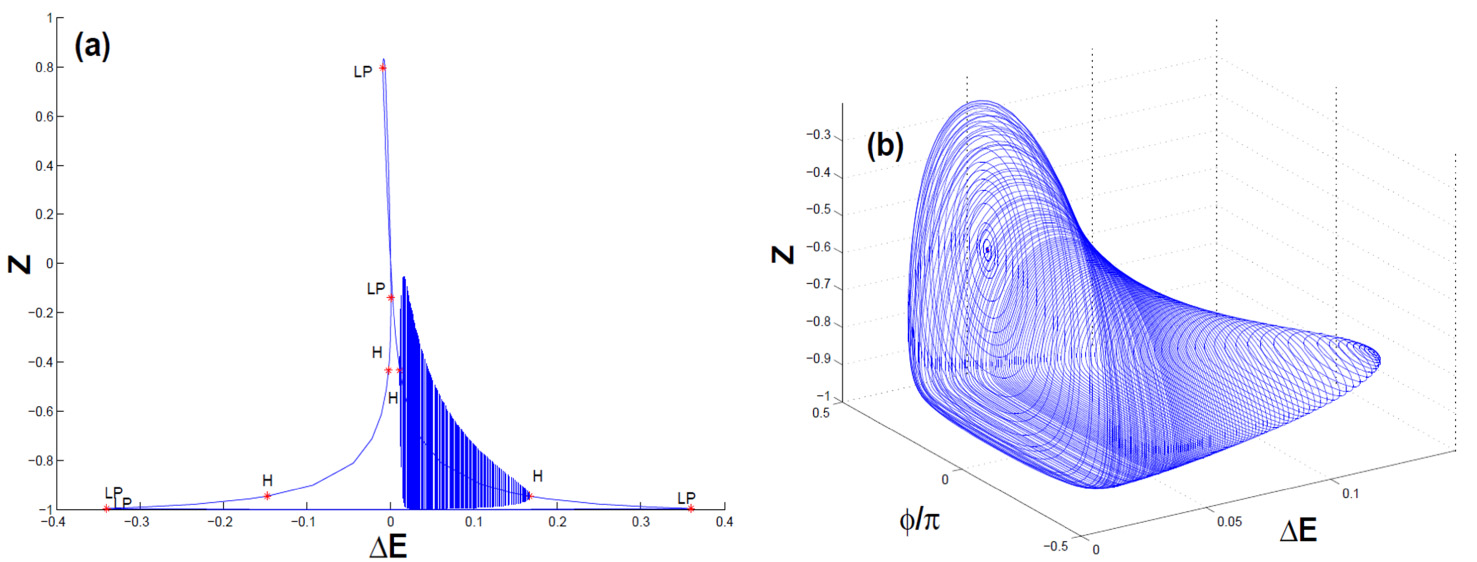,width=0.8\linewidth,clip=}
\caption{\scriptsize (Colour online) (a)Continuation of the stable point as a function of $\Delta E$. Model parameters are $kd = 6, \Lambda = 0.01$, $\zeta = 0.0095$. Supercritical limit cycles are spawned between the two Hopf-bifurcation points, (b)Visualization of the limit cycles indicated in Fig.~\ref{fig:6}(a) in the $Z-\phi -\Delta E$ parametric phase-space.}
\label{fig:6}
\end{figure*}

Next, the continuation is performed in the asymmetry parameter $\Delta E$. Figures~\ref{fig:6}(a) and~\ref{fig:6}(b) show these results for $\zeta = 0.0095$. In Table~\ref{table:DeltaE_bifurcations}, the bifurcation points are characterized and the respective normal form coefficients are listed. The first Lyapunov coefficient is negative for the Andronov-Hopf points which indicates supercritical bifucation. For clarity, the limit cycles are plotted only on the $\Delta E>0$ side; they are spawn between the two Andronov-Hopf points on the $\Delta E>0$ side, too.

\begin{table}[ht]
\caption{Bifurcation points shown in Fig.~\ref{fig:6}(a)}
\centering % used for centering table
\begin{tabular}{l cccc} % centered columns
\hline\hline %inserts double horizontal lines
Label & $\Delta E$ & $\phi^{*}$ & $Z^{*}$ & $a$\\ [0.5ex] % inserts table heading
\hline % inserts single horizontal line
LP & $-0.340$ & $0.878$ & $-0.997$ & $3.917$\\
H & $-0.148$ & $0.984$ & $-0.946$ & $-0.00626$\\
H & $-0.00257$ & $0.990$ & $-0.434$ & $-0.00275$\\
LP & $0.00319$ & $-1.01$ & $-0.139$ & $0.219$\\
LP & $0.00909$ & $-0.311$ & $0.797$ & $-0.0148$\\
H & $0.0112$ & $0.0102$ & $-0.4336$ & $-0.00098$\\
H & $0.167$ & $0.0165$ & $-0.9461$ & $-0.0623$\\
LP & $0.360$ & $0.0102$ & $-0.434$ & $-3.92$\\[1ex] % [1ex] adds vertical space
\hline %inserts single line
\end{tabular}\\
{LP:Limit Point (fold), H:Hopf point, a is the normal form coefficient for LP and first Lyapunov coefficient for H.}
\label{table:DeltaE_bifurcations} % is used to refer this table in the text
\end{table}

\begin{figure}[ht!]
\epsfig{file=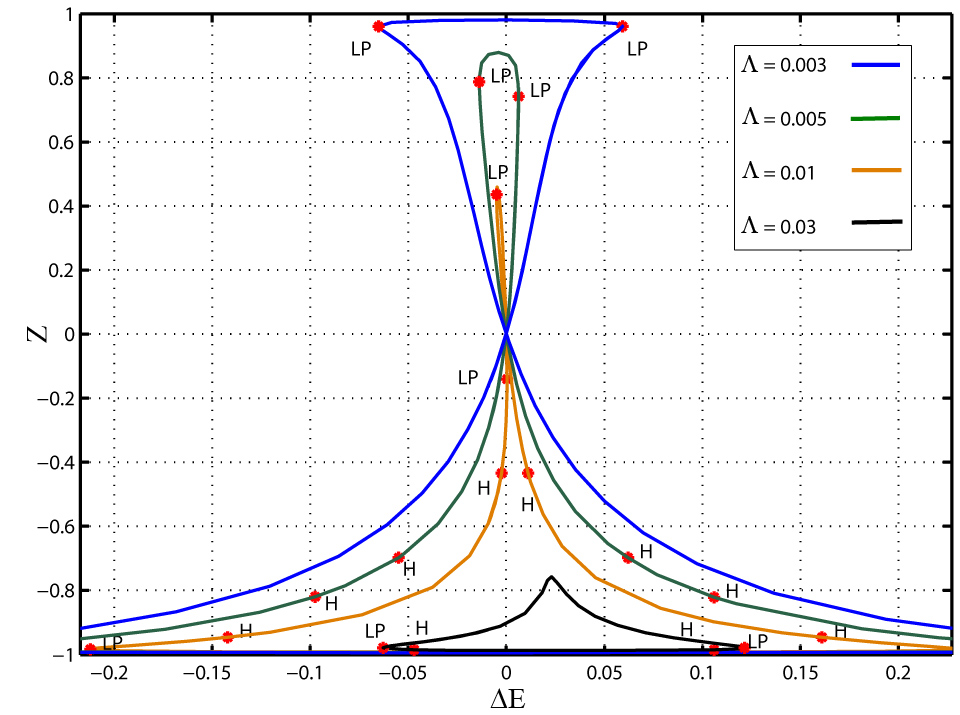,width=0.8\linewidth,clip=}
\caption{\scriptsize (Colour online) $\Delta E$ continuation of the stable point for different values of $\Lambda$, with $kd = 6$, $\zeta = 0.005$. Hopf (H) and limit point bifurcations are labeled on the continuation curves.}
\label{fig:7}
\end{figure}

\begin{figure}[ht!]
\epsfig{file=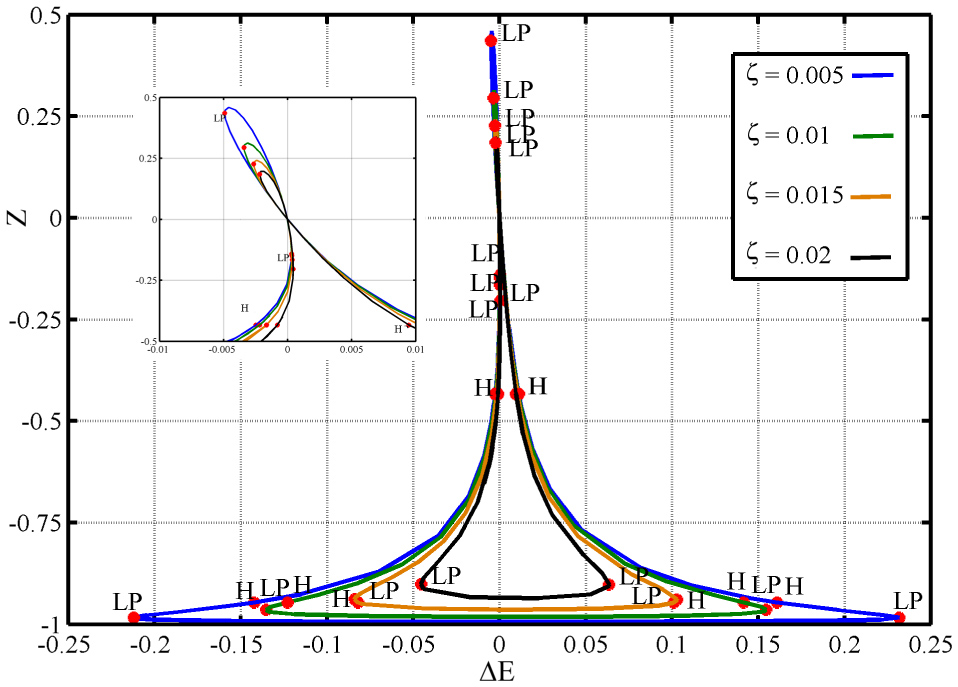,width=0.8\linewidth,clip=}
\caption{\scriptsize (Colour online) $\Delta E$ continuation of the stable point for different values of $\zeta$, with $kd = 6$, $\Lambda = 0.01$. Hopf (H) and limit point bifurcations are labeled on the continuation curves. Inset shows enlargement around the $\Delta E = 0, Z = 0$ region. Blue curve corresponds to the yellow curve in Fig.~\ref{fig:7}}
\label{fig:8}
\end{figure}

In order to gain further insight about the bifurcation, we investigate the continuation curve in the $Z-\Delta E$ plane for different values of $\Lambda$ in Fig.~\ref{fig:7} and for $\zeta$ in Fig.~\ref{fig:8}. Equilibrium points are periodic in the $Z-\phi$ plane, thus, each continuation curve closes onto itself in these parametric planes. The continuation curves extend in a finite region of $\Delta E - Z$ plane, where the range also depends on other model parameters. 

In Figure~\ref{fig:7}, we depict the effect of nonlinearity ($\Lambda$) on the continuation curve. In view of Eq.~\ref{eq:q} we note that, for a given soliton amplitude, increasing nonlinearity confines the soliton more tight around its propagation axis and decreases the coupling between the soliton and the SP. Thus, while the continuation curve extends in the whole $Z$ axis for small $\Lambda$ (blue curve) it shrinks and gets confined to the range close to $Z = -1$ for strong nonlinearity (black curve). Note that multiple equilibria are present for a given $\Delta E$ value when $|\Delta E|$ is small (e.g. $<$ $0.05$). In addition, the nonlinearity also induces different bifurcation points in the continuation curve. For $\Lambda = 0.003$ only LP occur. For $\Lambda = 0.03$ LP and H bifurcations are present but the first Lyapunov exponent is positive (i.e. subcritical Hopf bifurcations). For the intermediate values between these extremes, we observe that the continuation curve hosts multiple H-points which appear in pairs in negative and positive sides of $\Delta E$ axis. These are supercritical, and as it was shown in Fig.~\ref{fig:6} stable limit cycles are spawn between them.

When the dissipation strength $\zeta$ increases, the continuation curve in the $\Delta E - Z$ plane shrinks as shown in Fig.~\ref{fig:8}. Limit point and Hopf bifurcations decorate the continuation curves. The Hopf points close to $Z = -1$ are neutral saddle points. The other Hopf points are supercritical. As it was shown in Fig.~\ref{fig:5} stable limit cycles bifurcate from these Hopf points. Although not shown here, the curve completely dissappears for $\zeta > 0.028$ since no critical point is present.
  
We recall that $\Delta E = \Lambda - \nu_p$ is the difference between the (soliton-characteristic) nonlinearity and the (SP-characteristic) propagation constant. While not obvious in these terms, it is analogous to the asymmetry between the wells of the double well potential in BJJ systems, and thus, sizes the asymmetry between the soliton state and the SP state of the Josephson junction. In Figures~\ref{fig:7} and~\ref{fig:8}, we observe that for large magnitudes of $\Delta E$ (i.e. irrespective of its sign), the equilibrium points lie in the SP-dominant ($-1< Z << 0$) region. Soliton dominant equilibrium points occur under weak nonlinearity and weak dissipative conditions only.

\subsection{Codimension-2 Bifurcations}
\begin{figure}[ht]
\epsfig{file=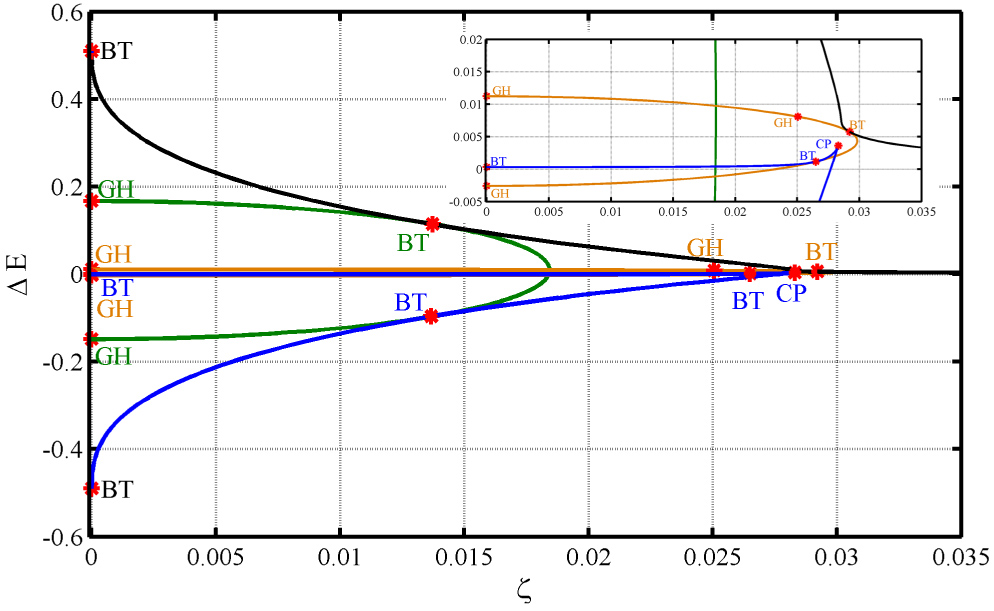,width=0.8\linewidth,clip=}
\caption{\scriptsize (Colour online) Codimension-2 bifurcation diagram in the $\zeta - \Delta E$ plane. Model parameters are $kd= 6, \Lambda = 0.01$. Blue curve shows the continuation of an equilibrium point. Orange and green curves are the Hopf continuation curves from corresponding BT points. Black curve is the limit point continuation curve. BT: Bogdanov-Takens point, C: Cusp point. GH: Generalized Hopf point. Inset shows the zoomed view around the $\Delta E=0$ axis.}
\label{fig:9}
\end{figure}

Codimension-2 bifurcations require the simultaneous variation of two parameters. The blue curve in Fig.~\ref{fig:9} shows the continuation an equilibrium point in the $\zeta-\Delta E$ parameter space. We observe three Bogdanov-Takens (BT) bifurcation points and a cusp point. BT-bifurcation implies the existence of three codimension-one bifurcations nearby, namely, a saddle-node bifurcation, an Andronov-Hopf bifurcation and a homoclinic bifurcation. From the BT points, the continuation curves of the Hopf point and limit point can be obtained respectively. The orange curve is the Hopf curve passing through the BT point close to the cusp point. The green curve is the Hopf curve passing through the lower BT point. Each Hopf curve has two generalized Hopf-points when they intersect the $\zeta = 0$ axis. The black curve is the continuation of the limit point from the upper BT point on the green Hopf curve.  

This codimension-2 diagram indicates the regions of $\zeta$ and $\Delta E$ in which stable limit cycles can be generated in the system. In Figure~\ref{fig:6} it was shown that for fixed $\zeta$ stable limit cycles are spawn between two Hopf points. Here, this corresponds to the region between the orange and the green curves. The Cusp point on the blue curve marks the value of $\zeta = 0.028$ beyond which there are no critical points for $\Lambda = 0.1$ and $\Delta E = -0.0025$ as it was shown in Fig.~\ref{fig:2}(b) and Fig.\ref{fig:4}(b). In Table~\ref{table:codim2_bifurcations}, the coordinates of the bifurcation points on the blue curve are given in the $\zeta - \Delta E$ plane and $\phi - Z$ plane and their normal form coefficients are listed. BT points have two normal form coefficients with respect to two varying parameters. Cusp point is characterized by a single normal form coefficient.

\begin{table*}[ht!]
\caption{Codimension-2 bifurcations in the $\zeta - \Delta E$ parameter space (blue curve in Fig~\ref{fig:9})}
\centering % used for centering table
\begin{tabular}{c c c c c c} % centered columns
\hline\hline %inserts double horizontal lines
Label & $\zeta$ & $\Delta E$ & $\phi^{*}$ & $Z^{*}$ & $(a,b)^{(1)}, c^{(2)}$\\ [0.5ex] % inserts table heading
\hline % inserts single horizontal line
BT & $0$ & $3.19\times10^{-4}$ & $-1.00$ & $-0.139$ & $(1.51\times10^{-4},-1.46\times10^{-7})$\\
BT & $0.026$ & $1.16\times10^{-3}$ & $-1.35$ & $-0.434$ & $(1.38\times10^{-4},-7.18\times10^{-2})$\\
CP & $0.028$ & $3.61\times10^{-3}$ & $-1.46$ & $-0.65$ & $-0.10$\\
BT & $0.014$ & $-0.096$ & $-1.27$ & $-0.946$ & $(-0.12,0.29)$\\ [1ex] % [1ex] adds vertical space
\hline %inserts single line
BT: Bogdanov-Takens Point, CP:Cusp point.\\
$^{(1)}$ normal form coefficients of BT.
$^{(2)}$ normal form coefficient of CP.
\end{tabular}
\label{table:codim2_bifurcations} % is used to refer this table in the text
\end{table*}

%\newpage
\section{Conclusion}
\label{sec:Conclusion}
In this work, we investigated the dynamical features of a dissipative nonlinear Josephson junction which is formed by a paraxial optical soliton surface-plasmon coupled system. The most notable part of the  heuristic model implemented here is the nonlinear nature of the coupling mechanism. For a dissipation proportional to the angular velocity, the system decays into stable fixed points with constant relative phase and constant average population imbalance. A particular feature observed here is the phase-slip phenomenon, where the odd-$\pi$ modes of the dissipationless system decay into even-$\pi$ modes. For dissipation proportional to the population imbalance, supercritical Andronov-Hopf bifurcation can occur which is notable in that stable oscillatory modes between the soliton and surface-plasmon fields can be maintained against strong dissipation. The heavily damped limit resembles the behavior of Superconducting Josephson junctions in which the system decays rapidly into a unique oscillating state. In view of this analysis, we propose that the nonlinear Josephson junction between the optical soliton and surface plasmons can exhibit interesting dynamical features which may be exploited further based on more rigorous theoretical models.
%\newpage
\begin{acknowledgments}
We acknowledge the support by the Science and Technology Research Council of Turkey (TUBITAK) under the project no: 111T285. K. G\"{u}ven acknowledges the support by the Turkish Academy of Sciences.
\end{acknowledgments}

\end{document}